# Electric Field Driven Torque in ATP Synthase

John H. Miller, Jr.,[1]* Kimal I. Rajapakshe,[1,2] Hans L. Infante,[1] and James R. Claycomb[1,3]

[1] Department of Physics and Texas Center for Superconductivity, University of Houston, Houston, Texas, USA.
[2] Department of Molecular and Cellular Biology, Baylor College of Medicine, Houston, Texas, USA.
[3] Department of Mathematics and Physics, Houston Baptist University, Houston, Texas, USA.

* Corresponding author, e-mail: jhmiller@uh.edu

## Abstract

$F_O$-ATP synthase ($F_O$) is a rotary motor that converts potential energy from ions, usually protons, moving from high- to low-potential sides of a membrane into torque and rotary motion. Here we propose a mechanism whereby electric fields emanating from the proton entry and exit channels act on asymmetric charge distributions in the *c*-ring, due to protonated and deprotonated sites, and drive it to rotate. The model predicts a scaling between time-averaged torque and proton motive force, which can be hindered by mutations that adversely affect the channels. The torque created by the *c*-ring of $F_O$ drives the $\gamma$-subunit to rotate within the ATP-producing complex ($F_1$) overcoming, with the aid of thermal fluctuations, an opposing torque that rises and falls with angular position. Using the analogy with thermal Brownian motion of a particle in a tilted washboard potential, we compute ATP production rates *vs.* proton motive force. The latter shows a minimum, needed to drive ATP production, which scales inversely with the number of proton binding sites on the *c*-ring.

## Introduction

Living organisms rely on molecular machines and pumps [1] that transport ions through membranes, enable movement, and carry out myriads of other activities. Such machines usually extract their energy from adenosine triphosphate (ATP), consisting of adenosine bound to three phosphate groups. Ion-driven rotary motors, by contrast, are driven by an electric potential and ion gradient across a membrane, somewhat like a battery-powered electric motor. There are three known types of rotary electric motors/pumps in biology [2]: the $F_O$-portion ($F_O$) of ATP synthase (F-ATPase, the smallest known rotary electric motor), the bacterial flagellar motor, and the $V_0$-portion of V-type ATPases. They contain a rotor (*c*-, *MS*-, or *K*-ring) embedded in a membrane. The rotor can be driven to rotate by ions moving from high to low potential or, in some cases and especially in V-type ATPases, by extracting energy from ATP hydrolysis as they pump protons or other ions "uphill" against the concentration gradient and/or membrane potential.

During ATP synthesis, the ion motive force across a mitochondrial, bacterial, or thylakoid membrane drives the *c*-ring of $F_O$ ATP synthase to rotate [3-5]. Protons, or occasionally sodium ions, enter a half-channel in the stator (*a*-subunit [6]) and are directed onto aspartate or glutamate binding sites (*c*Asp-61 for *E. coli*, glutamate for mitochondria) on the *c*-ring (Fig. 1). After completing nearly one revolution, each ion is expelled from its respective subunit into the exit half-channel coupling to the low-potential side of the membrane in Fig. 1. The *c*-ring drives the *γ*-subunit to rotate like a camshaft within $F_1$, structurally supported by the *b*-subunit [7], as it extracts and releases three ATP molecules per cycle from their catalytically active sites after phosphorylating ADP. Unless inhibited, for example by the inhibitor protein $IF_1$ [8], $F_OF_1$ ATP synthase can also use the energy from ATP hydrolysis in $F_1$ to drive $F_O$ to pump protons "uphill" against the membrane potential and concentration gradient.

The model proposed here provides an intuitively clear mechanism for torque generation and rotation in the *c*-ring of $F_O$. It is partly motivated by the fact that, in biology, electromagnetism plays the dominant role among the four forces of nature. Previous models of F-type ATPases [9-13] rely on complex computer simulations that fail to provide a clear picture of the torque-generation mechanism. Kaim and Dimroth [14] find that, in mitochondria, the membrane potential itself drives $F_O$ even in the absence of a significant proton gradient, suggesting that the electric field emanating from the channels plays a key role. ATP synthase is highly relevant to medicine due to its crucial bioenergetic role in humans, and its dysfunction has been implicated in cancer [15-19], neurodegenerative diseases [20,21], and other illnesses. The discussions below will thus treat the ions binding to the *c*-ring as protons, which are relevant to mitochondrial ATP synthase in humans and other eukaryotic organisms.

## Results and Discussion

The membrane-bound portion of ATP synthase ($F_O$) consists of the *a*-subunit containing the proton channels, the rotating *c*-ring, and the *b*-subunit that provides structural support. The torque generation mechanism proposed here is based on the turnstile rotary mechanism proposed by Vik and Antonio [3], similar to Junge *et al*'s model [4,22], in which the *a*-subunit contains two offset half-channels allowing entry and exit of protons that bind to the rotor (Fig. 1). Protons enter one half-channel in the *a*-subunit of the $ab_2$ stator on the high-potential side of the membrane (top of Fig. 1), bind to the rotating *c*-ring (e.g., $c_{10}$ rotor for *E. coli*), and exit the half-channel on the low-potential side of the membrane (bottom of Fig. 1).

The two offset proton channels couple to high- and low-potential sides of the mitochondrial or bacterial inner membrane. The high-potential side also often has a higher proton concentration, resulting in a pH gradient that contributes to the overall proton motive force (pmf). The proton channels are water-accessible pathways [23-25] that act as proton selective water wires [26]. Because of this proton selectivity, even when the membrane potential is reduced by compensating ion gradients ($Na^+$, $K^+$, $Cl^-$, etc.), the average *effective* potential difference between the termini of the proton half-channels is expected to be comparable to the total pmf or difference in proton chemical potentials, $\Delta p$. This is similar to the measured voltage across an electronic device (e.g., diode, tunnel junction, or battery) being the difference in *electron* chemical potentials ($\approx$ Fermi energies for metals or semiconductors). The imbalance of protons and potentials between the proton half-channels creates an electric dipole moment that generates electric field lines going around the *c*-ring and between channels, as shown in Fig. 2. The field emanating from the half-channels can only generate a net torque if, on average over time, one or more proton binding sites on the *c*-ring immediately between the half-channels are *deprotonated* while most of the remaining sites going around the *c*-ring are *protonated*. Without this charge imbalance the torques would cancel out.

Figure 2 shows cross-sections of the resulting equipotential surfaces (black lines) and tangential electric field components (red arrows), superimposed on an idealized cross section of the *c*-ring (yellow) and *a*-subunit (green). The proton channel cross sections in the *a*-subunit are colored to depict the differences in potentials, with dark blue and red representing the channels coupled to high- and low-potential sides of the membrane, respectively. The protonated sites on the *c*-ring are shown as dark blue circles, while the light circle represents a deprotonated site. The black arrows represent tangential forces due to the field acting on protonated and deprotonated sites. Crucially, both can make positive contributions to the torque since opposite field directions are counterbalanced by opposite charges. Moreover, while the *a*Arg-210 residue (*E. coli* numbering) on the *a*–subunit creates a barrier to positive charge that encourages deprotonation of the *c*-ring, it acts as a narrow potential well to negative charge, allowing ballistic transport of valence electrons that screen the deprotonated site as the *c*-ring rotates.

The orientation of Fig. 1, with the high-potential side on top, enables one to envision a gravity-driven mechanical model of $F_O$-ATP synthase, as shown in Fig. 3. One or more balls ("protons") enter a channel, roll down a spiral ramp like the walkway in New York City's Guggenheim museum [27] while pushing against the "*c*-ring," and emerge from an exit channel. The downward slope of the ramp depicts the negative gradient of electric potential (electric field) between the high- and low-potential channels. The model makes the direction of rotation intuitively obvious. It also shows that torque is maximized provided, most of the time on average, any sites between the channels are deprotonated while most sites going around the *c*-ring are protonated. In this picture, improperly protonating a site immediately between channels would be equivalent to trying to lift a ball up a much steeper ramp, representing the large, oppositely directed electric field between channels. Clearly such a ball would prefer to roll backwards down the ramp and would thus contribute negatively to the torque. This intuitive picture of $F_O$ differs from models that rely primarily on Brownian ratchet-like fluctuations to drive *c*-ring torque, e.g. [28], although both pictures predict the same direction of rotation. Thermal fluctuations, however, are crucial to overcoming the rather large energy barriers to release ATP from $F_1$. Any small-step jerky behavior of $F_O$, resulting from the binding and unbinding of protons from the *c*-ring, will thus be dominated by the much larger 120° steps in the $F_OF_1$ motor. Our subsequent discussions, which include coupling to $F_1$, explicitly treat the role of thermal fluctuations in overcoming the energy barriers in $F_1$.

The gravity-driven model, unfortunately, cannot capture the effects of negative charge of the deprotonated site(s) between channels (white circle in Fig. 2). The deprotonated site(s) also likely play important, if not dominant, roles in generating torque since the negative charge couples to an enormous electric field going between the two half channels. Moreover, it is immaterial whether any excess positive and negative charges are concentrated on the protonated and deprotonated binding sites themselves or distributed more broadly in the *c*-ring. The crucial idea is that, on average, there should be a suitable electrostatic charge imbalance on the *c*-ring resulting, averaged over time, from protonation and deprotonation at appropriate locations. Disruptions of the proton entry and exit channels caused by mutations of the mitochondrial gene *ATP*6 in humans can shunt protons in a possible water wire leak between the channels and/or improperly protonate and deprotonate the *c*-ring. This can lead to sometimes devastating neurodegenerative and other diseases in which the ability to produce ATP is greatly impaired [15-21].

In order to compute the net torque from $F_O$, the residual charge of each protonated site on the *c*-ring of Fig. 2 is taken to be *αe*, where *α* is expected to range from *zero* to *one* and $e = 1.60 \times 10^{-19}$ C, while the charge of each deprotonated site is taken to be $-(1-\alpha)e$. As discussed in **Methods**, summing the torques due to field-induced forces, using a continuum approximation for the charge on the *c*-ring, the time-averaged torque *τ* is found independently of *α* to scale with Δ*p* as:

$$\tau = \frac{ne}{2\pi}\Delta p . \qquad (1)$$

Here *n* is the total number of proton binding sites on the *c*-ring. The work to complete one revolution, $W = 2\pi\tau$, equals the free energy $ne\Delta p$ given up by the protons as they move from high- to low-potential, so Eq. (1) represents an upper bound of 100% efficiency for *τ* vs. Δ*p*.

When the *c*-ring couples to $F_1$ via the *γ*-subunit, the $F_O$-generated torque works against an opposing torque and potential due to $F_1$, as illustrated in Fig. 4a, due to the energy needed to release ATP and overcome an additional energy barrier. Three ATP molecules are synthesized per rotation [29] as the *γ*-stalk rotates in 120º (2π/3 radian) steps within the *α-β* hexamer of $F_1$ (green & orange "lollipop" in Fig. 1). The free energy ΔG required to synthesize ATP is expected to lie in the range 34-57 kJ/mol (0.35-0.59 eV/molecule) for ATP/ADP ratios ranging from $10^{-1}$ to $10^3$ [30]. The ATP/ADP ratio is likely close to *one* within the catalytically active portion of the $F_1$ complex. Equating ΔG, depicted in bottom plot of Fig. 4(a), to release ATP from its catalytically active binding site in $F_1$ [31] to the work to drive the torque through a 120º (2π/3 radian) angle, $W = \tau\Delta\theta = \tau(2\pi/3)$, yields a critical torque $\tau_c$ needed to sustain ATP production in the range 27-45 pN·nm. This does not include any additional torque needed to overcome the energy barrier in $F_1$ shown in Fig. 4(a), which is typically aided by thermal fluctuations as will be discussed. Moreover, this barrier is much larger than the barrier for proton entry and exit from the *c*-ring and any step-like behavior of the motor is thus dominated by the 120° steps of $F_1$ [32,33], showing Brownian ratchet-like behavior.

The above values of $\tau_c$ are consistent with measurements of torque on the *γ*-subunit [32,33], produced by $F_1$ when hydrolyzing ATP or needed to drive $F_1$ in the synthesis direction, e.g. using magnetic nanorods [32]. These show average torque values in the range ~30-45 pN·nm, with periodic (vs. angle) maximum values of ~50-60 pN·nm to overcome the energy barrier in the synthesis direction, and higher maximum values (~70-80 pN·nm), in the absence of ADP and phosphate, to overcome the larger energy barrier while driving rotation via *γ*-subunit slip [32]. Combining ΔG = (2π/3)$\tau_c$ with Eq. (1) yields a critical pmf – the minimum needed to drive ATP synthesis at finite temperatures – given by:

$$\Delta p_c = \frac{3\Delta G}{ne}, \qquad (2)$$

which also results from energy conservation. In the case of mammalian mitochondria, it has recently been found that the *c*-ring has only 8 *c*-ring subunits and proton (glutamate) binding sites [34], as compared to *n* = 10 for *E. coli* and yeast mitochondria [35]. Taking *n* = 8 and ΔG = 0.45 eV yields Δ$p_c$ = 169 mV. If Δ*p* falls below Δ$p_c$, oxidative ATP production comes to halt.

Torque measurements on $F_1$ show that its energy landscape is a periodic [32,33] function of the $\gamma$-subunit rotational angle $\theta$, with a dominant period of 120º or $2\pi/3$ radians. In our calculations of ensemble-averaged rotation and ATP production rates for $F_OF_1$ we assume a sinusoidal energy landscape (Fig. 4(a)), with an opposing torque from $F_1$ of the form: $\tau_c + \tau_1 \sin\phi$, where $\phi = 3\theta$. Here $\tau_c$ is the critical torque needed to release ATP from the $F_1$ binding site with the aid of thermal fluctuations, and $\tau_1$ results from the washboard potential barrier in Fig. 4(a) revealed by torque measurements [32]. Dissipative effects and thermal fluctuations can be modeled using a Langevin equation equivalent to Brownian motion of a particle in a tilted washboard potential. When $\tau > \tau_c$ (top of Fig. 4(a)), thermally activated hopping enables the system to overcome the barrier at a rate that increases with temperature. Moreover, the highly viscous environment in which ATP synthase is embedded replaces any inertial response $I\,\partial^2\phi/\partial t^2$ ($I$ being a scaled moment of inertia) with a damped response, $\eta\,\partial\phi/\partial t$, where $\eta$ is a damping coefficient. The result is the following Langevin equation [36]:

$$\eta\dot\phi = \tau - \tau_c - \tau_1 \sin\phi + \sqrt{2\eta kT}\,\xi(t). \tag{3}$$

Here $k$ is Boltzmann's constant and $\xi(t)$ represents thermal fluctuations assumed to be Gaussian white noise (see **Methods**).

To compute the ensemble average rotation rate vs. torque, Eq. (3) is converted to an equation for a probability distribution $P(\phi,t)$, which is solved numerically, as discussed in **Methods**. Since $\xi(t)$ is random, the time-averaged rotation rate obtained from Eq. (3) reduces to the following simple linear form when $\tau_1 = 0$:

$$\eta\langle\dot\phi\rangle = \tau - \tau_c. \tag{4}$$

Since $\phi = 3\theta$ and three ATP molecules are released per cycle in a properly functioning molecule, in the presence of substrate the time-averaged ATP production rate is given by:

$$\langle\dot{\mathrm{ATP}}\rangle = \frac{\langle\dot\phi\rangle}{2\pi}. \tag{5}$$

The average rotation and ATP production rates, in the presence of substrate, can thus be scaled to a dimensionless form using the following definition:

$$f \equiv \frac{\eta\langle\dot\phi\rangle}{\tau_c}. \tag{6}$$

All but two of the plots in the following discussions assume $\tau_c = 40$ pN·nm. Thus we use the definition: $f \equiv \eta\langle\dot\phi\rangle/\tau_{40}$, where $\tau_{40} \equiv 40$ pN·nm, in order to show apples-to-apples comparisons in our plots of rotation and ATP production rates. After the computations yield $f$ vs. torque $\tau$, the torque-pmf scaling relation, Eq. (1), is then used to obtain plots of normalized ensemble average rotation and ATP production rates vs. proton motive force $\Delta p$. Such ensemble averages are relevant to experiment, since actual measurements of ATP production and oxygen consumption rates in whole mitochondria reveal the average behavior of large numbers of ATP synthase molecules acting in parallel.

Figure 4(b) shows resulting plots of numerically computed $f$ vs. proton motive force $\Delta p$ for three different temperatures within the range for which water remains liquid: 2°C, 40°C, and 80°C. These plots are obtained using the values: $n = 8$, $\tau_c = 40$ pN·nm, and $\tau_1 = 20$ pN·nm. The inset shows plots of $f$ vs. temperature over the range 2-80°C for two different pmf's, 240 mV and 260 mV. Although thermal hopping over the barrier in Fig. 4(a) is expected to have a nonlinear thermally activated form vs. absolute temperature $T$ for values of $\Delta p$ slightly above $\Delta p_c$, the increase from 2°C to 80°C only represents a 28% increase in absolute temperature (275 - 353 K). Thus the rotation rate vs. temperature appears rather linear over this limited range, as the inset to Fig. 4(b) indicates.

One of the most dramatic effects of the scaling law in Eq. (1) is to cause a wide variation in the minimum $\Delta p$ needed to drive ATP production for different organisms, assuming the minimum energy and torque to release ATP remain roughly constant. Structural studies to date show numbers of $c$-ring subunits and ion (e.g., proton) binding sites ranging from $n = 8$ for mammalian mitochondria [34] to $n = 15$ for the ATP synthase $c$-ring in the photosynthetic thylakoid membrane of *Spirulina platensis* [37].

The effect of this variation in $n$ is shown in Fig. 5(a), which plots $f$ vs. $\Delta p$ for several values of $n$ ranging from 8 to 15. All of the plots assume a temperature of 37°C and values of $\tau_c = 40$ pN·nm and $\tau_1 = 20$ pN·nm. The negative values of $f$ represent backward rotation of the $c$-ring, where ATP hydrolysis in $F_1$ provides energy to pump the protons against the membrane potential and pH gradient. Such proton pumping processes are usually inhibited by $IF_1$ in ATP synthase motors of mammalian mitochondria [8], but likely play a crucial role in maintaining the needed membrane potentials of prokaryotic organisms. A key result depicted by Fig. 5(a) is the nearly factor of two variation in minimum $\Delta p$ needed to yield a positive $f$. From an evolutionary perspective, there may be a tradeoff between having a smaller minimum $\Delta p$ to drive oxidative ATP production (large $n$) vs. requiring fewer protons per ATP molecule (small $n$). Alternatively, perhaps there are selective advantages to the values of membrane potential maintained by different organisms and organelles for their specific environments. Finally, some variations in $n$ and $\Delta p$ could merely be evolutionary accidents to which various types of organisms and organelles have managed to adapt.

Since there may also be some variation in the minimum energy and energy barrier to release ATP, caused by slight structural variations in $F_1$ and other factors, we have also computed $f$ vs. $\Delta p$ for fixed $n$ (=8) and several values of $\tau_1$ and $\tau_c$, keeping the temperature fixed at 37°C, as shown in Figs. 5(b) and 5(c). Fig. 5(b) shows that increasing $\tau_1$, proportional to the potential energy barrier in Fig. 4(a), increases the spread in $\Delta p$ between ATP synthesis and ATP hydrolysis. By contrast, changing $\tau_c$, proportional to the free energy $\Delta G$ to phosphorylate and release ATP, simply shifts the overall $f$ vs. $\Delta p$ curve to the left or right, as illustrated by Fig. 5(c).

Nicholls discusses the decline in mitochondrial membrane potential that humans and other mammals typically experience as we age, and how it affects ATP production [38]. Using his reported values of maximum ATP/ADP ratio vs. proton motive force $\Delta p$ obtained from rat liver mitochondria [38], we find that the ATP production rate computed using our model fits the reported data at room temperature (24°C) using $n = 8$, $\tau_c = 31$ pN·nm, and $\tau_1 = 17$ pN·nm, as shown in Fig. 5(d). One usually measures the membrane potential $\Delta\psi$ using a voltage sensitive dye such as TMRM or TMRE [39] rather than the total proton motive force $\Delta p$. Unlike thylakoid membranes in chloroplasts where the pH gradient usually dominates $\Delta p$ [40], the membrane potential $\Delta\psi$ often dominates $\Delta p$ ($\Delta\psi > 70\%$ of $\Delta p$) in mitochondria [30].

The need to maintain adequate $\Delta p$ to produce ATP via oxidative phosphorylation has enormous medical consequences. Inadequacy of mitochondrial membrane potential and pH gradient likely play important roles in the latter stages of a failing heart [41] and other problems resulting from accumulated mitochondrial mutations due to aging [42]. Mitochondrial complex I deficiency [43] is a frequently reported cause or contributor to cardiomyopathies and neurodegenerative disorders. Dysfunction of complex I and other mitochondrial complexes upstream from ATP synthase, and their inability to maintain adequate $\Delta p$, almost certainly results in inadequate ATP production. The simple ATP synthase model discussed here could thus provide a conceptual framework to study a range of effects of mitochondrial dysfunction, by incorporating it into a more comprehensive model of the respiratory chain.

## Methods

### A. *Electric Field Driven Torque in $F_O$-ATP Synthase*

The equipotential surface cross-sections (curved lines) in Fig. 2 were calculated using QuickField finite-element method software. Figure 6 shows a simplified geometry illustrating the proton channel and *c*-ring cross sections. Following our discussion in the previous section regarding the proton selectivity of the two channels then, if one labels the two channel positions as *A* and *B*, the electric field inside the membrane is expected to satisfy: $\int_A^B \mathbf{E} \cdot d\mathbf{r} = -\Delta p$, where $\Delta p$ is the proton motive force. Though not essential for subsequent derivations, one can obtain a convenient approximate analytical expression for the field in the middle of the membrane by treating the channels (cross sections shown in Fig. 6) as long cylinders of radius *a* with centers separated by distance *d*, charge per unit volume $\pm\rho$ yielding charge per unit length $\pm\lambda = \pm\pi a^2 \rho$, and assuming $a \ll d \ll r$, where *r* is the radius of the c-ring. A rather lengthy calculation (e.g., see [44]), taking the vector sum of the electric fields, then yields:
$E_\theta = (\lambda d / 8\pi\varepsilon r^2)[\tan^2(\theta/2) + 1]$, where $\varepsilon$ is the dielectric constant and $\theta$ is defined in Fig. 6. Here we hypothesize that the pmf $\Delta p$ across the membrane (see discussion of proton selectivity in Results and Discussion) is equivalent to the effective potential difference $\Delta V$ between channels. Thus, by equating $\Delta V$ to the negative line integral of the field between channels yields: $\Delta p = \Delta V = \lambda \ln 4 / (2\pi\varepsilon) \Rightarrow \lambda = 2\pi\varepsilon\Delta p / \ln 4$. Substituting this result into expression for $E_\theta$ above yields the following simple expression for the tangential field component:

$$E_\theta = E_0 \left[ \tan^2 \frac{\theta}{2} + 1 \right] = E_0 \left[ \tan^2 \delta + 1 \right], \qquad \text{(M-1)}$$

where the angles are defined in Fig. 6 and $E_0 = \Delta p \, d / (4 \ln 4 \, r^2)$.

At this stage we take the number of normally occupied ion (usually proton) binding sites in the *c*-ring to be *p*, the number of normally unoccupied sites to be *q*, and the total number of ion binding sites to be $n = p + q$. In the case of $F_0$, *q* is usually ~1-2, while *n* ranges from 8 to 15 [34,37]. The tangential force acting on each occupied site (*p*-site) will then be $F_{p\theta} = \alpha e E_\theta$, where $e = 1.60 \times 10^{-19}$ C and $\alpha$ represents the fraction of uncompensated positive charge, where $0 \leq \alpha \leq 1$. Similarly, the tangential force acting on each unoccupied site ($q$-site) will be $F_{q\theta} = -(1-\alpha)eE_\theta$. In order to compute time-averaged torque (smoothing out stepping motion), we make a continuum approximation, where any residual positive charge in the occupied sites is distributed uniformly around the $c$-ring perimeter, and similarly for any net negative charge in the unoccupied site(s). Additionally, we assume that the half channels are (optimally) positioned such that $d/2\pi r = q/n$. Within these approximations, the average charge per unit length outside (in the $p$-sites) and between ($q$-sites) half channels are given by $\rho_p = n\alpha e/(2\pi r)$ and $\rho_q = -n(1-\alpha)e/(2\pi r)$. Finally, we note that the proton motive force satisfies:

$$\Delta p = \int_{\substack{\text{outside} \\ \text{path}}} \mathbf{E}_{\text{out}} \cdot \mathbf{dr} = \int_{\substack{\text{inside} \\ \text{path}}} \mathbf{E}_{\text{in}} \cdot \mathbf{dr} \text{ , where the line integrals are from the positive to the negative}$$

half channel around the ring (outside path) or between half channels (inside path). The total field-induced torque, averaged over time, then becomes:

$$\tau = r \int_{\substack{\text{outside} \\ \text{path}}} \rho_p \mathbf{E}_{\text{out}} \cdot \mathbf{dr} - r \int_{\substack{\text{inside} \\ \text{path}}} \rho_q \mathbf{E}_{\text{in}} \cdot \mathbf{dr} = \frac{rne}{2\pi r}\left[\alpha \Delta p + (1-\alpha)\Delta p\right]. \quad \text{(M-2)}$$

Eq. (M-2) does not depend on the detailed form of the electric field (e.g., Eq. (M-1) or not) provided we take $\rho_p$ and $\rho_q$ to be constant in the regions of interest. Doing the appropriate cancellations in Eq. (M-2) yields the scaling law, Eq. (1), between torque and proton motive force, which is repeated here for convenience:

$$\tau = \frac{ne}{2\pi}\Delta p \ . \quad \text{(M-3)}$$

Here $n$ is the total number of ion (proton) binding sites on the rotor. The work to complete one revolution is $W = 2\pi\tau$. This, according to Eq. (M-3), is precisely the energy $ne\Delta p$ given up by the protons as they translocate across the membrane, so the scaling law represents the upper bound of 100% efficiency in converting pmf to torque. More generally, a scaling of the form $\tau = \beta\Delta p$, obtained either from molecular dynamics simulations or experiment, would yield a torque generation efficiency given by $\mathcal{E} = 2\pi\beta/ne$. Due to cancellations in Eq. (M-2), Eq. (M-3) applies regardless of the value of $\alpha$, i.e., whether the torque primarily comes from coupling of the field to the deprotonated site between half-channels or to any residual charge due to the protonated sites going around the $c$-ring. The key is that an overall charge asymmetry is needed regardless of the precise location of charge.

The dynamics of the rotor, in the absence of coupling of the $c$-ring to the ATP producing complex $F_1$, is readily modeled by assuming overdamped motion and taking $\eta'$ to be a viscous damping coefficient, leading to an equation of motion for the rotor-$\gamma\varepsilon$ assembly: $\eta'\dot\theta = \tau = (ne/2\pi)\Delta p$. The ion (eg. proton) current $J$ through the $c$-ring is the total transported charge times the number of revolutions per unit time:

$$J = ne\frac{\dot\theta}{2\pi} = G_0\Delta p, \quad \text{(M-4)}$$

where $G_0 = n^2 e^2 / [4\pi^2 \eta']$ is the conductance of F$_O$ assuming any central hole in the *c*-ring is plugged by the $\gamma\varepsilon$ subunit. Equation (M-4) is consistent with the experimentally observed ohmic conductance [45] in the proton-driven rotor of F$_O$-ATP synthase in the absence of F$_1$.

If the average occupancy $p_o/p$ of ideally occupied *p*-sites is less than *one*, and similarly with the number $q_e$ of empty *q*-site(s), then the torque will be reduced, as follows:

$$\tau = \frac{ne}{2\pi}\left[\frac{p_o}{p} + \frac{q_e}{q} - 1\right]\Delta p, \tag{M-5}$$

which simplifies to Eq. (M-3) (Eq. (1)) when $p_o/p = 1$ and $q_e/q = 1$. One can see from the above equation, Eq. (M-5), that improper protonation and deprotonation of the *c*-ring due to mutations or other factors will reduce the torque vs. proton motive force efficiency.

B. *Coupling to the F$_1$-ATP Synthase Energy Landscape*

When F$_O$ is coupled to F$_1$ via the $\gamma$-subunit, the torque generated by F$_O$ works against an opposing torque and potential due to F$_1$, as illustrated in Fig. 4(a). Three ATP molecules are synthesized per rotation [29] as the $\gamma$-stalk rotates in 120° steps within the $\alpha$-$\beta$ hexamer of F$_1$ (green & orange "lollipop" in Fig. 1). Equating the free energy $\Delta G$ to release ATP from its binding site in F$_1$ [31] to the work, $W = \tau \Delta\theta = \tau(2\pi/3)$, done by the driving torque through a 120° ($2\pi/3$ radian) angle, yields a minimum torque $\tau_c = 3\Delta G/2\pi$ to sustain ATP production. Measurements of torque [32,33], produced by F$_1$ when hydrolyzing ATP or required to drive F$_1$ to phosphorylate ADP, yield values in the ranges discussed in the previous section. Combining $\Delta G = \tau_c (2\pi/3)$ with equation (M-3) yields a critical proton motive force – the minimum needed to drive ATP synthesis at finite temperatures – given by:

$$\Delta p_c = \frac{3\Delta G}{ne}, \tag{M-6}$$

which also results from energy conservation. Taking $n = 10$ (e.g., for *E. coli*) and $\Delta G = 0.45$ eV, yields $\Delta p_c = 135$ mV. Importantly, equations (M-3) and (M-6) universally apply to all three domains of life. If $\Delta p$ falls below $\Delta p_c$, ATP production via oxidative phosphorylation comes to halt.

Torque measurements [32,33] on F$_1$ also show that its energy landscape is periodic, with a dominant periodicity of 120°. When coupling F$_0$ to F$_1$, we assume a sinusoidal energy landscape (Fig. 4(a)), with an opposing torque of the form: $\tau_c + \tau_1 \sin 3\theta$. The equation of motion for the rotor-$\gamma\varepsilon$ assembly then becomes: $\eta'\dot{\theta} = \tau - \tau_c - \tau_1 \sin 3\theta = (ne/2\pi)\Delta\mu - \tau_c - \tau_1 \sin 3\theta$. At finite temperatures, allowing thermally activated hopping over the barrier, the critical proton motive force is simply: $\Delta p_c = 2\pi\tau_c/ne$, where $\tau_c = (3/2\pi)\Delta G$. Making the substitutions, $\phi = 3\theta$ and $\eta = \eta'/3$, leads to the following equation of motion:

$$\eta \frac{d\phi}{dt} = \tau - \tau_c - \tau_1 \sin\phi, \tag{M-7}$$

which describes a particle in a tilted washboard potential. At finite temperatures, Brownian fluctuations can be modeled using a Langevin equation [36]:

$$\eta \frac{d\phi}{dt} = \tau - \tau_c - \tau_1 \sin\phi + \sqrt{2\eta kT}\, \xi(t) \qquad \text{(M-8)}$$

where $k$ is Boltzmann's constant and $\xi(t)$ is assumed to be Gaussian white noise, $\langle \xi(t)\xi(t_1)\rangle = \delta(t - t_1)$. Equation (M-8) is equivalent to Brownian motion of a particle in a tilted washboard potential, for which a kinetic theory has been developed in the context of condensed matter systems [46-48].

At finite temperatures, following Ambegaokar and Halperin [46] (also see [49]), the Fokker-Planck equation resulting from equation (M-8) reduces in the overdamped limit to the Smoluchowski equation for a distribution function $P(\phi,t)$:

$$\frac{\partial P}{\partial t} = -\frac{1}{\eta}\frac{\partial}{\partial \phi}\left[(\tau - \tau_c - \tau_1 \sin\phi)P - kT\frac{\partial P}{\partial \phi}\right] \equiv -\frac{\partial W}{\partial \phi}. \qquad \text{(M-9)}$$

In Eq. (M-9), the flux $W = \langle \dot\phi\rangle/2\pi$ is equivalent to the time-averaged ATP production rate. The steady state case, $\partial P/\partial t = 0$, leads to the following first-order differential equation:

$$\frac{dP}{d\phi} + \frac{1}{kT}[\tau - \tau_c - \tau_1 \sin\phi]P = \frac{\eta}{kT}W, \qquad \text{(M-10)}$$

where $W$ is now constant. The rotation rate $\langle \dot\phi\rangle = 2\pi W$ is determined by finding the solution $P(\phi)$, and applying the normalization, $\int_{-\pi}^{\pi} P(\phi) = 1$, and periodicity condition, $P(-\pi) = P(\pi)$, with the result [38] (defining $\tau' \equiv \tau - \tau_c$):

$$\langle \dot\phi\rangle = \frac{2\pi kT}{\eta}\left[1 - \exp\left(-\frac{2\pi\tau'}{kT}\right)\right]\left[\int_0^{2\pi} F(x,0)\left(\int_x^{x+2\pi} F(0,y)dy\right)dx\right]^{-1}, \qquad \text{(M-11)}$$

where

$$F(k,l) = \exp\left[-\frac{1}{kT}\int_k^l [\tau' - \tau_1 \sin z]dz\right]. \qquad \text{(M-12)}$$

The steady state flux in equations (M-11) and (M-12) is a solution of the 1-D Smoluchowski equation for cyclic diffusion paths reported by Stratonovich [50] and used to characterize diffusion in tilted periodic potentials [51]. When $kT$ and $\tau'$ are both small compared to $\tau_1$, the above integrals reduce to the following analytical expression [52,53]:

$$\langle \dot{\phi} \rangle = \frac{2}{\eta} \sqrt{\tau_1^2 - \tau'^2} \exp\left\{ -\frac{\sqrt{\tau_1^2 - \tau'^2}}{kT} - \frac{\tau'}{kT} \sin^{-1} \frac{\tau'}{\tau_1} \right\} \sinh\left[ \frac{\pi \tau'}{2kT} \right]. \tag{M-13}$$

The results discussed in the previous section are obtained by numerically integrating Eqs. (M-11) and (M-12), enabling computation of rotation and ATP production rates vs. torque. The scaling relation between torque and $\Delta p$ (Eq. (1) or Eq. (M-3)) then enables plots of ATP production rate vs. proton motive force $\Delta p$, with finite-temperature results shown in Figs. 4(b) and 5.

## Concluding Remarks

The field-driven torque model proposed here suggests a simple mechanism by which a membrane potential and ion gradient can drive a rotary motor such as ATP synthase. When $F_O$-ATP synthase is coupled to $F_1$, the model shows that ATP synthase exhibits highly nonlinear behavior, with a critical proton motive force needed to drive ATP production that depends strongly on the number of proton binding sites on the *c*-ring. An increased, but not infinite, periodic potential barrier in $F_1$ in the absence of ADP [32] suggests that ATP synthase itself acts as a mitochondrial "uncoupler" when the proton motive force is high. Proton flux could drive the *c*-ring and *γ*-subunit of $F_OF_1$-ATP synthase to rotate even without producing ATP [32], thus dissipating some of the excess membrane potential that would otherwise produce deleterious reactive oxygen species when the mitochondrial membrane potential is sufficiently high.

In humans, congenital or somatic mitochondrial mutations in the gene *ATP*6, causing deleterious amino acid replacements in or near the *a*-subunit proton channels, may cause proton leaks and/or improper protonation of the *c*-ring. Such dysfunction could contribute to diseases that include congenital neurodegenerative diseases (e.g. Leigh's syndrome [54-56]), cardiomyopathies, and age-related diseases such as cancer [15-19,42,57]. The model discussed here provides an intuitive conceptual framework to study the effects of such mutations. Future work would potentially use more realistic models of ATP synthase based on structural data and employ molecular dynamics, adaptive Poisson-Boltzmann solvers, and other tools to study the effects of certain mitochondrial mutations on electrostatic potential distributions, proton translocation, and torque generation in $F_O$-ATP synthase.

## Acknowledgments

The authors gratefully acknowledge permission from R. H. Fillingame for use of an image adapted and modified to Fig. 1 and assistance in preparing figures from Sladjana Maric and Rooplekha Mitra.

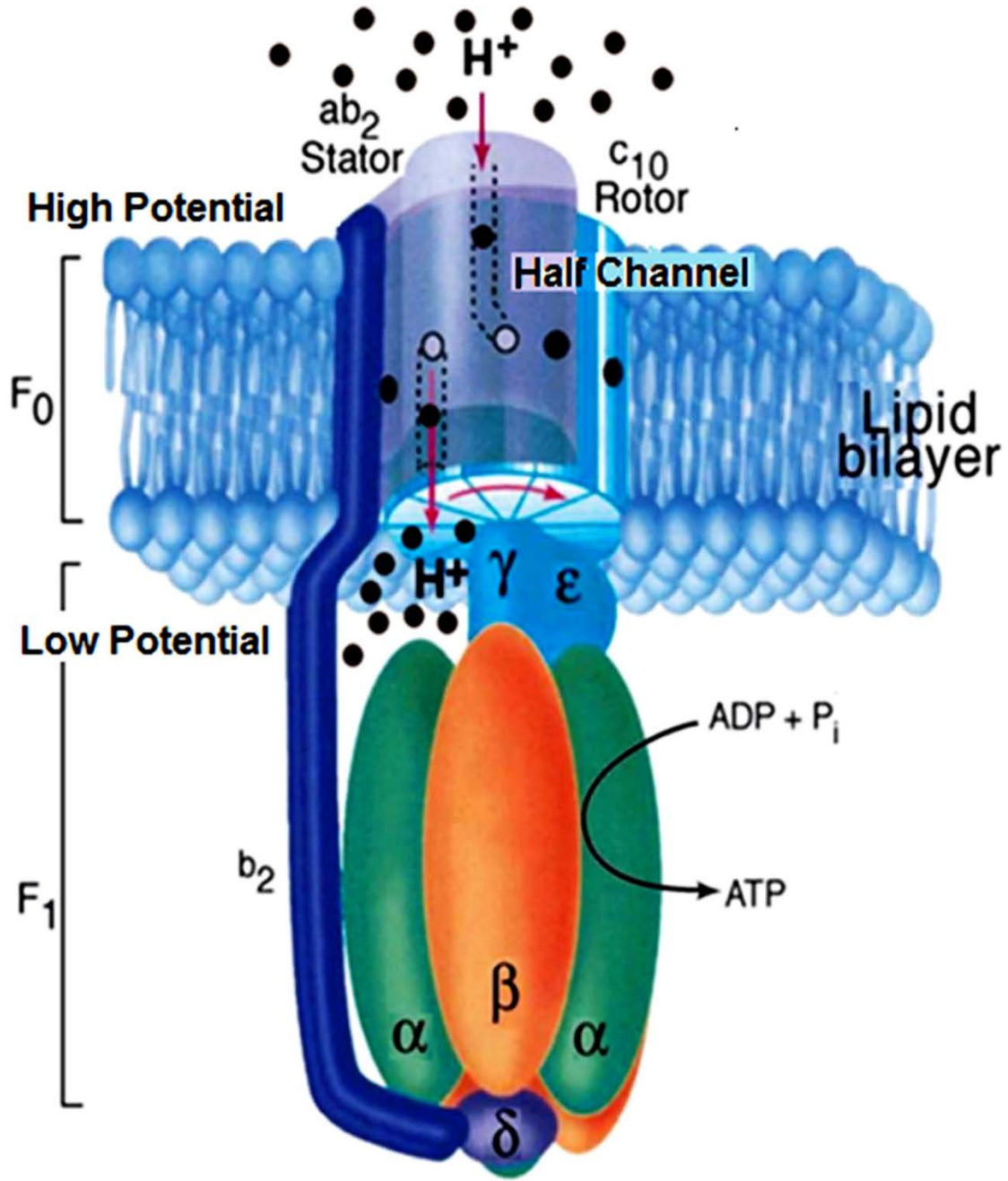

**Figure 1. Side view of $F_1F_O$ ATP synthase, oriented with high-potential side on top, showing the half-channels in the stator *a*-subunit, *c*-ring rotor of $F_O$ ($c_{10}$ for *E. coli*), and $F_1$ complex, within which the $\gamma$-subunit rotates to release three ATP molecules per cycle** [adapted from Fillingame [5] with permission].

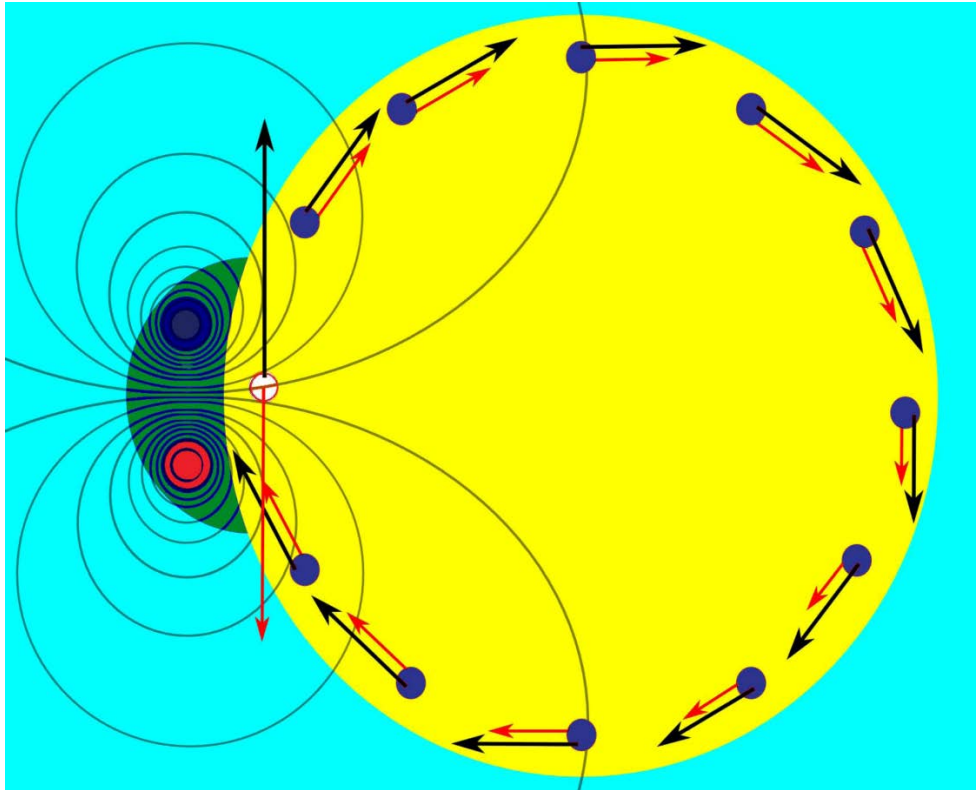

**Figure 2. Top view of the *c*-ring (yellow) and stator *a*-subunit (green) of F$_O$, showing equipotential surface cross-sections (curved lines) perpendicular to the electric field emanating from the half-channels (blue and red circles) in the *a*-subunit.** Black arrows represent forces due to tangential field components (red arrows) acting on protonated (blue circles) and deprotonated (light circle) sites on the *c*-ring. (Equipotentials computed using QuickField.)

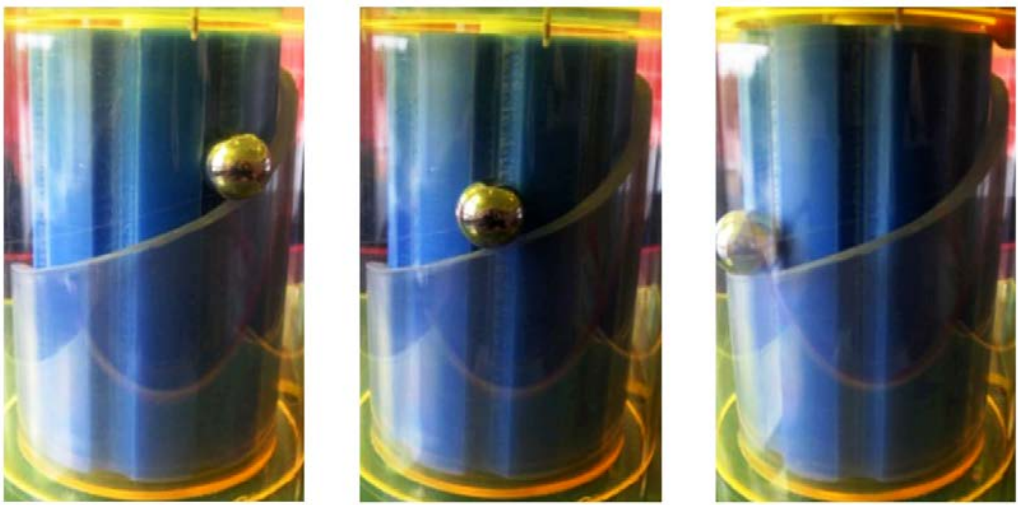

**Figure 3. Gravity-driven mechanical model of F$_O$, where the ball rolling down the ramp depicts a proton moving to lower potential energy, driving the blue "*c*-ring" to rotate, before emerging from the "exit channel" hole.**

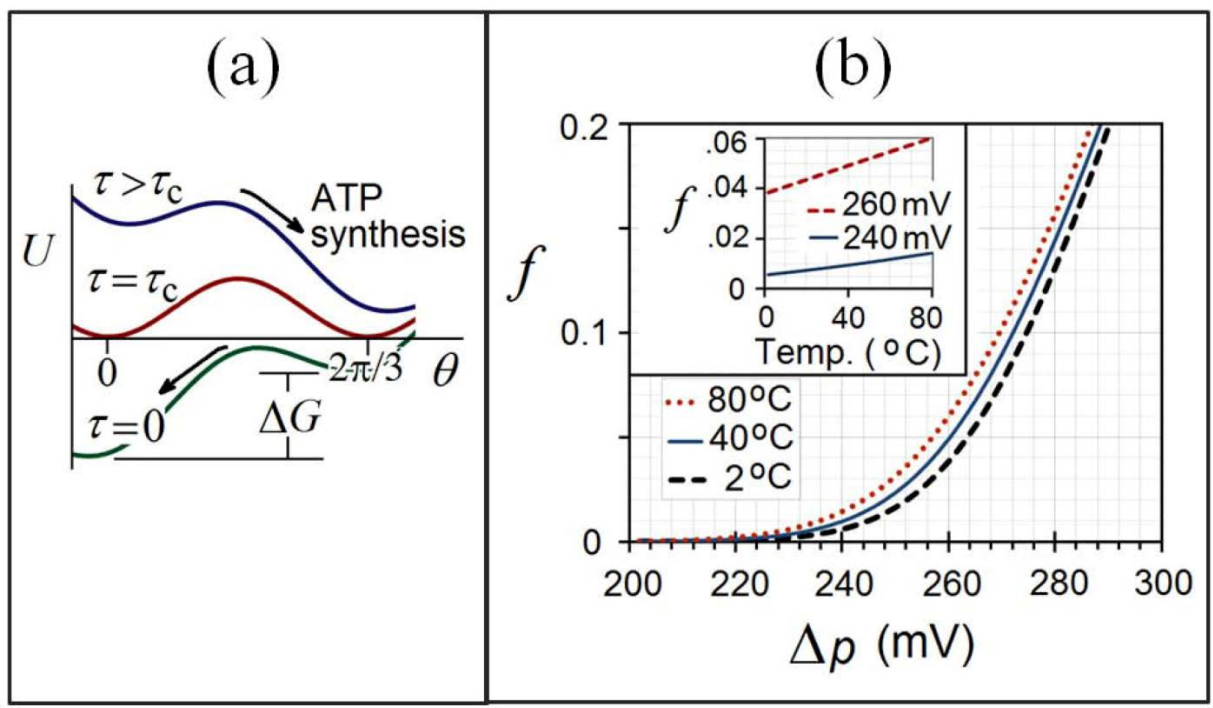

**Figure 4. Torque from the *c*-ring drives rotation and ATP production by overcoming the periodic energy barrier in $F_1$ with the aid of thermal fluctuations.**
**(a)** Tilted washboard potential $U(\theta)$ as the driving torque $\tau$ from $F_O$ works against the opposing torque $\tau_c + \tau_1 \sin(\theta/3)$ from $F_1$. **Top**: Torque exceeds the critical value ($\tau > \tau_c$) needed to release ATP from $F_1$ with the aid of thermally activated hopping.
**Bottom**: Plot shows the case where $F_1$ uses energy from ATP hydrolysis to drive the *c*-ring backwards and pump protons across the membrane. **(b)** Computed thermally assisted ATP production (rotation) rates $f$ vs. pmf $\Delta p$, using the tilted washboard potential in Fig. 4(a), at various temperatures and (**inset**) $f$ vs. temperature for two different pmf's, using the values $n = 8$, $\tau_c = 40$ pN·nm, and $\tau_1 = 20$ pN·nm.

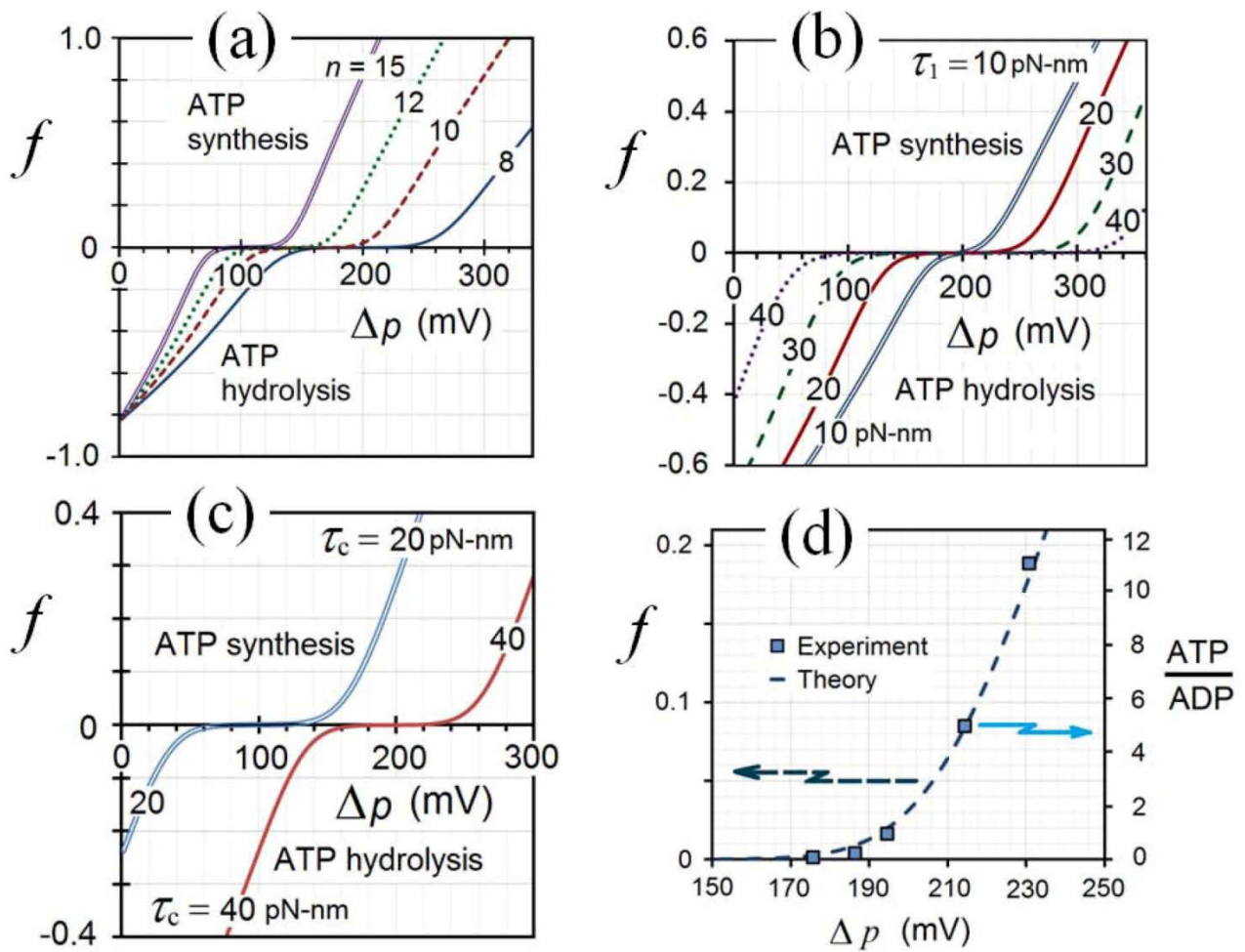

**Figure 5. Plots show the predicted *c*-ring rotation rates vs. proton motive force for various values of $n$, $\tau_c$, and $\tau_1$. (a)** Predicted rotation rate $f$ (see text) vs. pmf, $\Delta p$, taking $\tau_c = 40$ pN·nm and $\tau_1 = 20$ pN·nm, for various numbers $n$ of proton binding sites on the *c*-ring. Positive and negative rotation rates correspond to ATP synthesis and ATP hydrolysis, respectively. **(b)** Theoretical rotation rate $f$ vs. $\Delta p$, assuming $\tau_c = 40$ pN·nm and $n = 8$, for various values of $\tau_1$. **(c)** Predicted rotation rate $f$ vs. $\Delta p$, assuming $\tau_1 = 20$ pN·nm and $n = 8$, for two values of $\tau_c$. **(d)** Theoretical ATP production rate (dashed line) vs. $\Delta p$ (using $\tau_c = 31$ pN·nm and $\tau_1 = 17$ pN·nm), as compared to maximum intracellular ATP/ADP ratios (squares) reported by Nicholls [38].

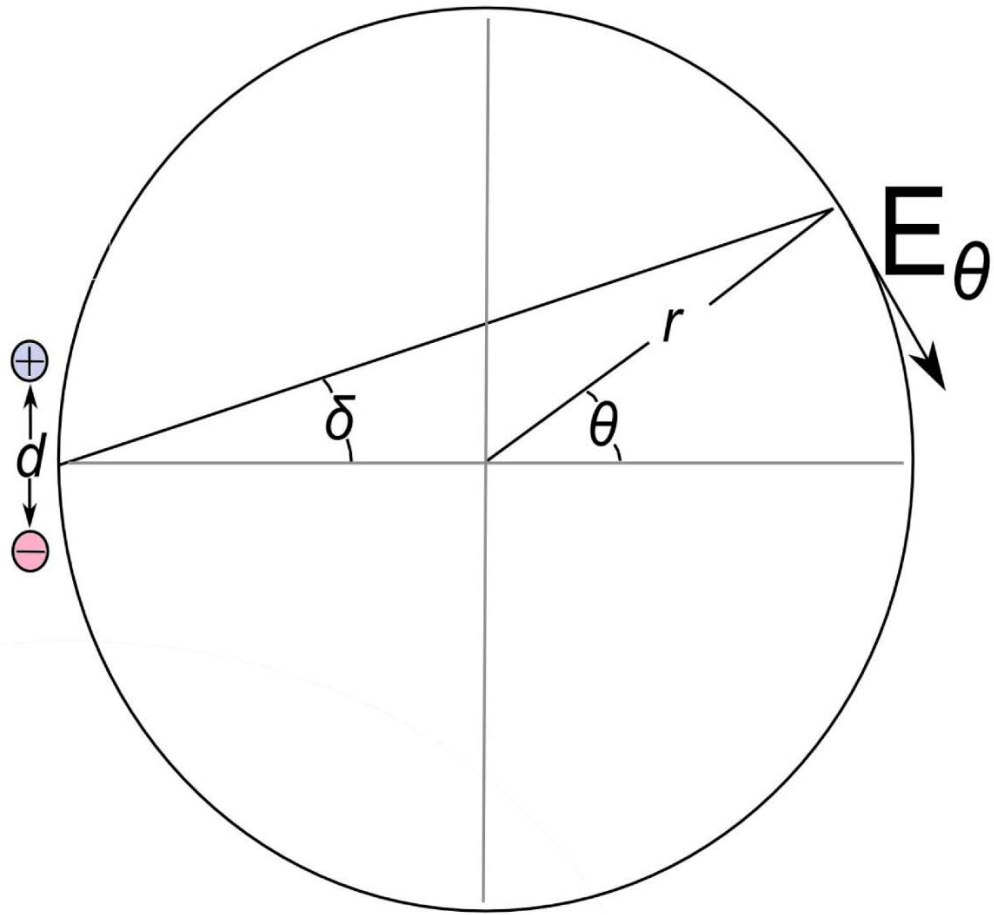

**Figure 6.** Geometry of *c*-ring and stator half channels, showing the geometrical parameters used to express the tangential field component in Eq. (M-1).